\documentclass[a4paper]{article}

\usepackage{icrc2013}
\usepackage[english]{babel}

\title{Development of the Photomultiplier-Tube Readout System for the CTA Large Size Telescope}

\shorttitle{Development of the PMT Readout System for the CTA LST}

\authors{
H.\,Kubo$^1$,
R.\,Paoletti$^2$,
Y.\,Awane$^1$,
A.\,Bamba$^3$,
M.\,Barcelo$^4$,
J.A.\,Barrio$^5$,
O.\,Blanch$^4$,
J.\,Boix$^4$,
C.\,Delgado$^6$,
D.\,Fink$^7$,
D.\,Gascon$^8$,
S.\,Gunji$^9$,
R.\,Hagiwara$^9$,
Y.\,Hanabata$^{10}$,
K.\,Hatanaka$^1$,
M.\,Hayashida$^{10}$,
M.\,Ikeno$^{11}$,
S.\,Kabuki$^{12}$,
H.\,Katagiri$^{13}$,
J.\,Kataoka$^{14}$,
Y.\,Konno$^1$,
S.\,Koyama$^{15}$,
T.\,Kishimoto$^1$,
J.\,Kushida$^{16}$,
G.\,Mart\'inez$^6$,
S.\,Masuda$^1$,
J.M.\,Miranda$^{17}$,
R.\,Mirzoyan$^7$,
T.\,Mizuno$^{18}$,
T.\,Nagayoshi$^{15}$,
D.\,Nakajima$^7$,
T.\,Nakamori$^9$,
H.\,Ohoka$^{10}$,
A.\,Okumura$^{19}$,
R.\,Orito$^{20}$,
T.\,Saito$^1$,
A.\,Sanuy$^8$,
H.\,Sasaki$^{21}$,
M.\,Sawada$^3$,
T.\,Schweizer$^7$,
R.\,Sugawara$^{20}$,
K.-H.\,Sulanke$^{22}$,
H.\,Tajima$^{19}$,
M.\,Tanaka$^{11}$,
S.\,Tanaka$^{13}$,
L.A.\,Tejedor$^5$,
Y.\,Terada$^{15}$,
M.\,Teshima$^{7,10}$,
F.\,Tokanai$^9$,
Y.\,Tsuchiya$^1$,
T.\,Uchida$^{11}$,
H.\,Ueno$^{15}$,
K.\,Umehara$^{13}$,
T.\,Yamamoto$^{21}$
for the CTA Consortium.
}

\afiliations{
$^1$Department of Physics, Graduate School of Science, Kyoto University, Sakyo, Kyoto 606-8502, Japan\\
$^2$Universit\`a di Siena, and INFN Pisa, I-53100 Siena, Italy\\
$^3$Department of Physics and Mathematics, Aoyama Gakuin University, Sagamihara, Kanagawa, 252-5258, Japan\\ 
$^4$Institut de F«¿sica d'Altes Energies, Edifici Cn, Campus UAB, 08193 Bellaterra, Spain\\
$^5$Grupo de Altas Energ«¿as, Universidad Complutense de Madrid, Av Complutense s/n, 28040 Madrid, Spain\\
$^6$CIEMAT, Avda. Complutense 22, 28040 Madrid, Spain\\
$^7$Max-Planck-Institut f\"{u}r Physik, F\"{o}hringer Ring 6, D 80805 M\"{u}nchen, Germany\\
$^8$Universitat de Barcelona (ICC-UB), Mart«¿ i Franqu«²s, 1, 08028, Barcelona, Spain\\
$^9$Department of Physics, Faculty of Science, Yamagata University, Yamagata, Yamagata 990-8560, Japan\\
$^{10}$Institute for Cosmic Ray Research, The University of Tokyo, Kashiwa, Chiba 277-8582, Japan\\
$^{11}$Institute of Particle and Nuclear Studies, KEK, Tsukuba, Ibaraki 305-0801, Japan\\
$^{12}$Department of Radiation Oncology, Tokai University, 143 Shimokasuya, Isehara-shi, Kanagawa 259-1193, Japan\\
$^{13}$College of Science, Ibaraki University, Mito, Ibaraki 310-8512, Japan\\
$^{14}$Research Institute for Science and Engineering, Waseda University, 3-4-1, Okubo, Shinjuku-ku, Tokyo, 169-8555, Japan\\
$^{15}$Graduate School of Science and Engineering, Saitama University, 255 Simo-Ohkubo, Sakura-ku, Saitama 338-8570, Japan\\
$^{16}$Department of Physics, Tokai University, 4-1-1 Kita-Kaname, Hiratsuka, Kanagawa 259-1292, Japan\\
$^{17}$Grupo de Electronica, Universidad Complutense de Madrid, Av. Complutense s/n, 28040 Madrid, Spain\\
$^{18}$Hiroshima Astrophysical Science Center, Hiroshima University, 1-3-1 Kagamiyama, Higashi-hiroshima, 739-8526 Japan\\
$^{19}$Solar-Terrestrial Environment Laboratory, Nagoya University, Chikusa, Nagoya 464-8601, Japan\\
$^{20}$Faculty of Integrated Arts and Sciences, The University of Tokushima, Tokushima 770-8502, Japan\\
$^{21}$Department of Physics, Konan University, 8-9-1 Okamoto Higashinada-ku Kobe, Hyogo, 658-8501 Japan\\
$^{22}$Deutsches Elektronen-Synchrotron, Platanenallee 6, 15738 Zeuthen, Germany\\
\scriptsize{
}
}

\email{kubo@cr.scphys.kyoto-u.ac.jp}

\abstract{
We have developed a prototype of the photomultiplier tube (PMT) 
readout system for the Cherenkov Telescope Array (CTA) Large Size 
Telescope (LST). Two thousand PMTs along with their readout systems 
are arranged on the focal plane of each telescope, with one readout 
system per 7-PMT cluster. The Cherenkov light pulses generated by the air showers 
are detected by the PMTs and amplified in a compact, low noise and wide dynamic 
range gain block. The output of this block is then digitized at a sampling 
rate of the order of GHz using the Domino Ring Sampler DRS4, an analog memory 
ASIC developed at Paul Scherrer Institute. The sampler has 1,024 capacitors 
per channel and four channels are cascaded for increased depth. After 
a trigger is generated in the system, the charges stored in the capacitors are 
digitized by an external slow sampling ADC and then transmitted via Gigabit Ethernet. 
An onboard FPGA controls the DRS4, trigger threshold, and Ethernet transfer. 
In addition, the control and monitoring of the Cockcroft-Walton circuit that 
provides high voltage for the 7-PMT cluster are performed 
by the same FPGA. A prototype named {\it Dragon} has been developed that 
has successfully sampled PMT signals at a rate of 2 GHz, and generated single 
photoelectron spectra.
}


\begin{document}
\maketitle

\section{Introduction}
A ground-based imaging atmospheric Cherenkov telescope (IACT) measures 
Cherenkov light from an extended air shower (EAS) generated by the 
interaction between very high energy (VHE) gamma rays and the upper 
atmosphere. Night sky background (NSB) also enters a pixel photon sensor 
of the focal plane of the IACT with a rate of the order of 10--100 MHz, 
depending on the mirror size and the pixel size. In some region of the 
Galactic plane, the NSB can reach on average up to several hundred MHz. 
The NSB therefore becomes noise that affects the sensitivity and the energy 
threshold. Given this NSB pollution, and the fact that the duration of 
Cherenkov light from EAS is a few nanoseconds, a fast digitization speed 
of the readout system coupled to a fast photosensor like a 
photomultiplier tube (PMT) is beneficial to increase the pixels' 
signal-to-noise ratio. In addition, this system should be compact and 
have low cost and low power consumption because each IACT possesses 
several thousand photon sensor pixels and the readout system attached to 
the sensors is in a \mbox{camera} container located at the focal position. 
Furthermore, a wide dynamic range of more than 8--10 bits is required 
to resolve a single photoelectron and have a wider energy range.

A commercial flash analog-to-digital converter (FADC) satisfies the
requirement of a wide dynamic range. However, it is costly and consumes
relatively high power of a few watts per channel. On the other hand, an
analog memory application specific integrated circuit (ASIC), that
\mbox{consists} of several hundred to several thousand switched capacitor
arrays (SCA) per channel, can sample a signal at the order of GHz, with
a wide dynamic range and lower power consumption. Several types of
analog memory ASICs have been developed for applications in particle
physics and cosmic ray physics. With respect to IACTs, a modified
version of ARS0 \cite{Lachartre}, Swift Analog Memory (SAM)
\cite{Delagnes}, and Domino Ring Sampler (DRS) \cite{Ritt} chips are
used in the H.E.S.S.-I, H.E.S.S.-II, and MAGIC experiments,
respectively.

Cherenkov Telescope Array (CTA) \cite{CTA} is the next generation VHE
gamma-ray observatory, which improves the sensitivity by a factor of 10
in the range 100 GeV--10 TeV and an extension to energies well below 100
GeV and above 100 TeV. CTA consists of telescopes having mirrors with
size 23 m, 12 m, 3.5--7 m, and $\sim$50 m$^2$, which are called large
size telescope (LST), medium size telescope (MST), small size telescope
(SST), and Schwarzschild-Couder telescope (SCT), respectively. Several
types of readout systems are developed for CTA with different analog
memories for the requirements of LST, MST, SST, and SCT. At the
same time, there has been progress in the development of photon sensors
such as PMT and a Geiger-mode avalanche photodiode (or silicon
photomultiplier) for CTA. At this time, the primary candidate for LST is a PMT.

 Using the analog memory DRS version 4 (hereafter DRS4), we have so far
 developed three versions of prototypes for the PMT readout system for
 CTA, named {\it Dragon}. Using the first version of the prototype, 
we demonstrated that the waveform of a PMT signal can be well
 digitized with the DRS4 chip. The second version was developed 
with improvements made to the sampling depth of DRS4 and 
a trigger \cite{icrc2011_kubo}.
The third version was developed to be downsized and reduce the cost for mass production. 
In this paper, we report the design and performance of the third version of the prototype.

\section{Design of Readout System}
\subsection{Overview}
Two thousand PMTs and their readout systems are arranged on 
the focal plane of each LST telescope \cite{Blanch}, with one readout system per 7-PMT
cluster. We have developed the third prototype of the PMT readout system. 
Figures 1 and 2 show a photograph and block diagram of the prototype, 
respectively. The prototype consists of a 7-PMT cluster, 
a slow control board and a DRS4 readout board. 
The total size is 14 cm $\times$ 48 cm. 
The 7-PMT cluster and the \mbox{slow} control board are described in 
detail in reference \cite{Orito}. In this paper, a brief description is provided. 

 Our 7-PMT cluster consists of seven head-on type \mbox{PMTs} with a
super-bialkali photocathode and 8-stage dynodes (Hamamatsu Photonics,
R11920-100 with a diameter of 38 mm) \cite{Toyama}, a Cockcroft-Walton (CW) circuit
for high voltage supply to the PMTs (designed by Hamamatsu Photonics),
and a preamplifier board. A signal from the PMT is amplified by a 
preamplifier, and fed to the DRS4 readout board. A commercial preamplifier, 
Mini-circuits LEE-39+, is used in the current {\it Dragon} (Fig. 3). 
A preamplifier board with an ASIC designed for CTA, PACTA \cite{PACTA}, 
has been recently developed and will be used in the next {\it Dragon}. 

 \begin{figure}[ht]
  \vspace{-3mm}
  \centering
  \includegraphics[width=7.7cm]{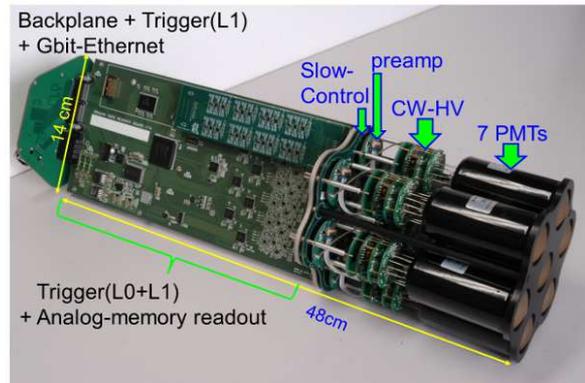}
  \vspace{-5mm}
  \caption{Photograph of the 7-PMT cluster and readout system (ver. 3).}
  \label{fig1}
 \end{figure}
 \begin{figure}[ht]
  \vspace{-7mm}
  \centering
  \includegraphics[width=7.8cm]{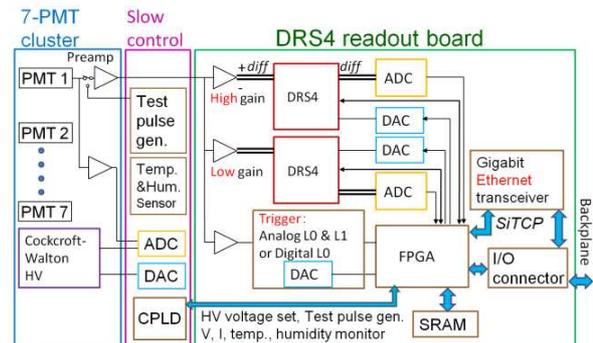}
  \vspace{-5mm}
  \caption{Block diagram of the 7-PMT cluster and readout system
  (ver. 3).}
  \label{fig2}
 \end{figure}

 On the DRS4 readout board the preamplified signal is divided into 
three lines: a high gain channel, a low gain channel, and a trigger
channel. The high and low gain \mbox{channels} are connected to DRS4 chips. 
The signal is sampled at a rate of the order of GHz and the waveform 
is stored in a \mbox{SCA} in DRS4. When a trigger is generated in the trigger 
circuit, the voltages stored in the capacitor array are sequentially 
output and then digitized by an external slow sampling ($\sim$30 MHz) 
ADC. The digitized data is sent to a field programmable gate array
(\mbox{FPGA}) and then transmitted to a Gigabit Ethernet transceiver and a 
backplane via a data input/output (I/O) connector. The \mbox{FPGA} controls 
a static random access memory (SRAM) that stores large amounts of data 
before transmission and a digital-to-analog converter (DAC) used for 
thresholding in the trigger circuit. 

 \begin{figure}[ht]
  \vspace{-2mm}
  \centering
  \includegraphics[width=7cm]{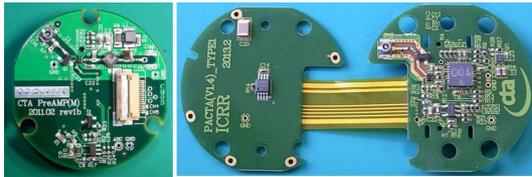}
  \vspace{-5mm}
  \caption{Preamplifier board with a commercial amplifier (left) and the PACTA (right).}
  \label{fig3}
 \end{figure}

 The slow control board (Fig.1) is equipped with a generator for test pulses that are fed to the preamplifier, a temperature and
humidity sensor with I2C interface, a DAC for setting the voltage of 
the CW high voltage circuit, and an ADC for monitoring both the CW 
circuit and the \mbox{DC} anode current. These devices on the slow control 
board are controlled by a complex programmable logic device (\mbox{CPLD}). 
Since the \mbox{CPLD} communicates with the \mbox{FPGA} on the DRS4 readout board, 
the data to and from the \mbox{CPLD} is sent via the Ethernet.
The power supply to the DRS4 readout board and the slow control board 
is $\pm$3.3V and +5V. The total power consumption is $\sim$2W per readout channel.

\subsection{DRS4 Readout Board}
Figure 4 shows a photograph of the DRS4 readout board with a size of
14 cm $\times$ 30 cm, 10 cm shorter than the version 2. 
The slow control board attached to the 7-PMT cluster is connected 
to the DRS4 readout board from the right side via 
two card-edge connectors. The DRS4 board has main amplifiers, 
eight DRS4 chips, ADCs for digitizing a signal stored in the capacitor array in DRS4 at a 
sampling frequency of 33 MHz, a DAC to control the DRS4, an \mbox{FPGA}, 
a 18Mbit SRAM, a Gigabit Ethernet transceiver, 
and a data I/O connector to the backplane. 
A Xilinx Vertex-4 \mbox{FPGA} was adopted in the version 2. In the version 3, 
a \mbox{Xilinx} Spartan-6 \mbox{FPGA} is used to reduce the cost for mass production. 
In addition, analog level 0 (L0) and 1 (L1) trigger mezzanines or 
a digital level 0 (L0) trigger mezzanine are mounted on the DRS4 board. 
Details of these parts are described in the following subsections.

 \begin{figure}[ht]
  \vspace{-4mm}
  \centering
  \includegraphics[width=8.4cm]{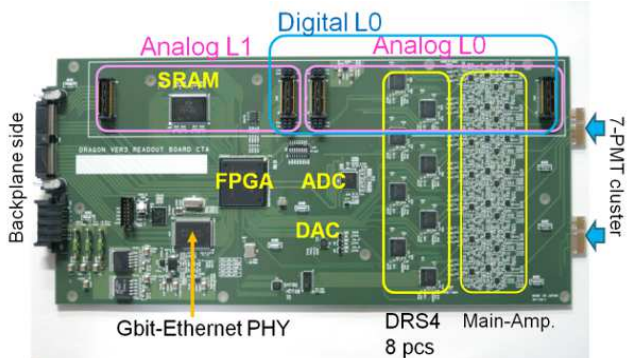}
  \vspace{-9mm}
  \caption{Photograph of the DRS4 readout board (ver. 3).}
  \label{fig4}
 \end{figure}

\subsection{Main Amplifier}
Figure 5 shows a block diagram of the main amplifiers. It is designed to have a bandwidth greater than 250 MHz for the high gain channel, lower power consumption, and a dynamic range of 0.2-1000 photoelectrons for LST. A preamplified signal from a PMT with a typical gain of 4 $\times$ 10$^4$ is fed to the main amplifier. The signal is amplified using \mbox{two} differential amplifiers to meet the requirement for bandwidth and power consumption. For the high gain channel, it is amplified by a gain of 9 using Analog Devices, ADA4927. For the trigger, it is amplified by a gain of 4 using Analog Devices, ADA4927 and ADA4950. For the low gain channel, the preamplified signal is attenuated by 1/4 and then buffered with a differential amplifier using Analog Devices, ADA4950.

 \begin{figure}[ht]
  \vspace{-2mm}
  \centering
  \includegraphics[width=5cm]{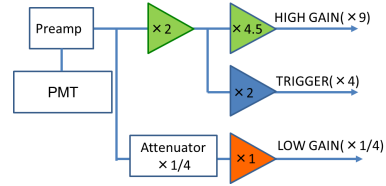}
  \vspace{-3mm}
  \caption{Block diagram of the main amplifiers.}
  \label{fig5}
 \end{figure}

\subsection{Trigger}
The DRS4 readout board has been designed to host an analog trigger \cite{Analog_Trig} 
or digital trigger \cite{icrc2011_kubo}. The different elements of the trigger system 
will be placed at different locations on the readout electronics, as shown in Fig. 4. 

For the analog trigger, the decision boards (L0/L1) are placed on the readout board itself, 
while for the digital trigger the L0 board is placed in the frontend and the L1 in the backplane, 
keeping both analog and digital L0 as close as possible after the signals from PMTs.
The distribution boards are placed at the backplane of the readout board, in order to send and receive 
trigger signals efficiently among neighbouring clusters. The choice of using trigger 
mezzanines and sharing of connectors has allowed the independent study of both analog and digital solution in the development phase.

\subsection{DRS4 Chip}
The DRS chip is being developed at the Paul Scherrer Institute (PSI),
Switzerland, for the MEG experiment \cite{Ritt}. 
The latest version, DRS4, is also used in the MAGIC experiment. 
The DRS4 chip includes nine differential input
channels at a sampling speed of 700 MS/s--5 GS/s, with a bandwidth of
950 MHz, and a low noise of 0.35 mV after offset correction. The analog
waveform is stored in 1,024 sampling capacitors per channel and the
waveform can be read out after sampling via a shift register that is
clocked at a maximum of 33 MHz for digitization using an external ADC. 
The power consumption of the DRS4 chip is 17.5 mW per channel at
2 GS/s sampling rate. The chip is fabricated using the 0.25 $\mu$m CMOS
technology.

 It is possible to cascade two or more channels to obtain deeper
 sampling depth. In the {\it Dragon}, four channels were
 cascaded, leading to a sampling depth of 4,096 for each PMT signal, 
which corresponds to a depth of 4 $\mu$s at 1 GS/s sampling rate. 
It enables continuous sampling until the stereo coincidence 
trigger confirms the local telescope trigger, and 
initializes the camera readout.

\subsection{FPGA-based Gigabit Ethernet (SiTCP)}
The digitized waveform data and the 
monitor/control data are transmitted via Ethernet with only two 
devices: FPGA and Gigabit Ethernet transceiver (PHY). This simple 
composition is available on a hardware-based TCP processor, 
SiTCP \cite{Uchida}. The circuit size of SiTCP in the FPGA is 
$\sim$3000 slices, which is small enough to allow implementation 
with user circuits on a single FPGA. In addition, the SiTCP 
has an advantage in that the throughputs in both directions 
can simultaneously reach the theoretical upper limits of Gigabit Ethernet.

\subsection{Mini-Camera with Three Clusters}
A prototype with three clusters of PMTs and their readout boards was constructed to develop the trigger system between multiple clusters (Fig. 6) \cite{Analog_Trig}.

 \begin{figure}[ht]
  \centering
  \includegraphics[width=8.4cm]{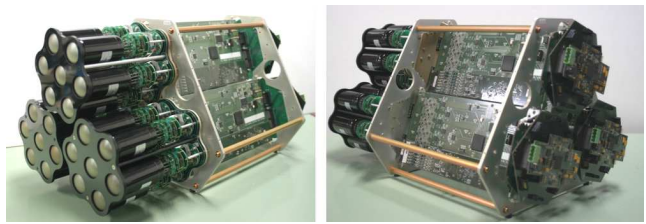}
  \vspace{-3mm}
  \caption{Prototype with three clusters of PMTs and their readout boards.}
  \label{fig6}
 \end{figure}

\section{Performance of Readout System}
The performance of the DRS4 readout system {\it Dragon} was measured. Figure 7 shows that dynamic ranges 
of high and low gain channels are 60 p.e. and 2500 p.e., which satisfied the requirement 
of $>$1000 photoelectrons. Measured noise level for the high gain channel is 0.1 p.e. (rms).

Figure 8 shows normalized gains for high and low gain channels as a function of frequency. 
The bandwidths for high and low gain channels are 260 MHz and 190 MHz 
at -3dB, respectively while those of the DRS4 readout board 
combined with a preamplifier (LEE-39+) were 250 MHz and 180 MHz 
at -3dB for high and low gain channels, respectively. 

After the above measurements, the performance of the DRS4 readout board attached to a PMT 
was investigated. Figure 9 shows a pulse shape of the high gain channel 
of the PMT signal with a gain of 5$\times$10$^4$, which was measured with 
a UV laser and the DRS4 readout system at a sampling rate of 2 GS/s. 
The PMT signal having a width of $\sim$3 ns (FWHM) and a height corresponding to 
$\sim$3 photoelectrons was successfully digitized. 
Measured dead time for readout of 60 cells in the DRS4 chip is 0.9\% and 5.4\% 
at 1 kHz and 7 kHz, respectively. Figure 10 shows a single photoelectron spectrum 
of the high gain channel of the PMT signal with a gain of 5$\times$10$^4$, 
which was measured with {\it Dragon} and an LED at a sampling rate of 2 GS/s. 
In the figure, a single photoelectron peak is clearly seen. 
The signal to noise ratio defined as (single-photoelectron mean - pedestal mean) / pedestal r.m.s. is 3.6.

 \begin{figure}[h]
  \vspace{-2mm}
  \centering
  \includegraphics[width=5.5cm]{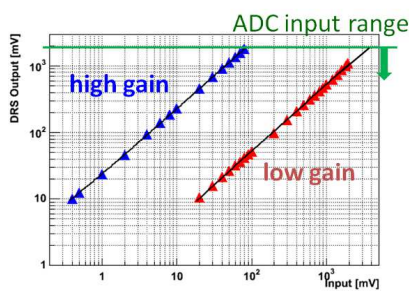}
  \vspace{-3mm}
  \caption{Dynamic ranges of the DRS4 readout board.}
  \label{fig7}
 \end{figure}
 \begin{figure}[h]
  \centering
  \includegraphics[width=5.5cm]{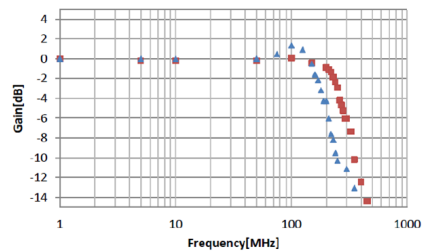}
  \vspace{-3mm}
  \caption{Bandwidths of high-(square) and low-(triangle) gain channels.}
  \label{fig8}
 \end{figure}
 \begin{figure}[h]
  \vspace{-2mm}
  \centering
  \includegraphics[width=5.5cm]{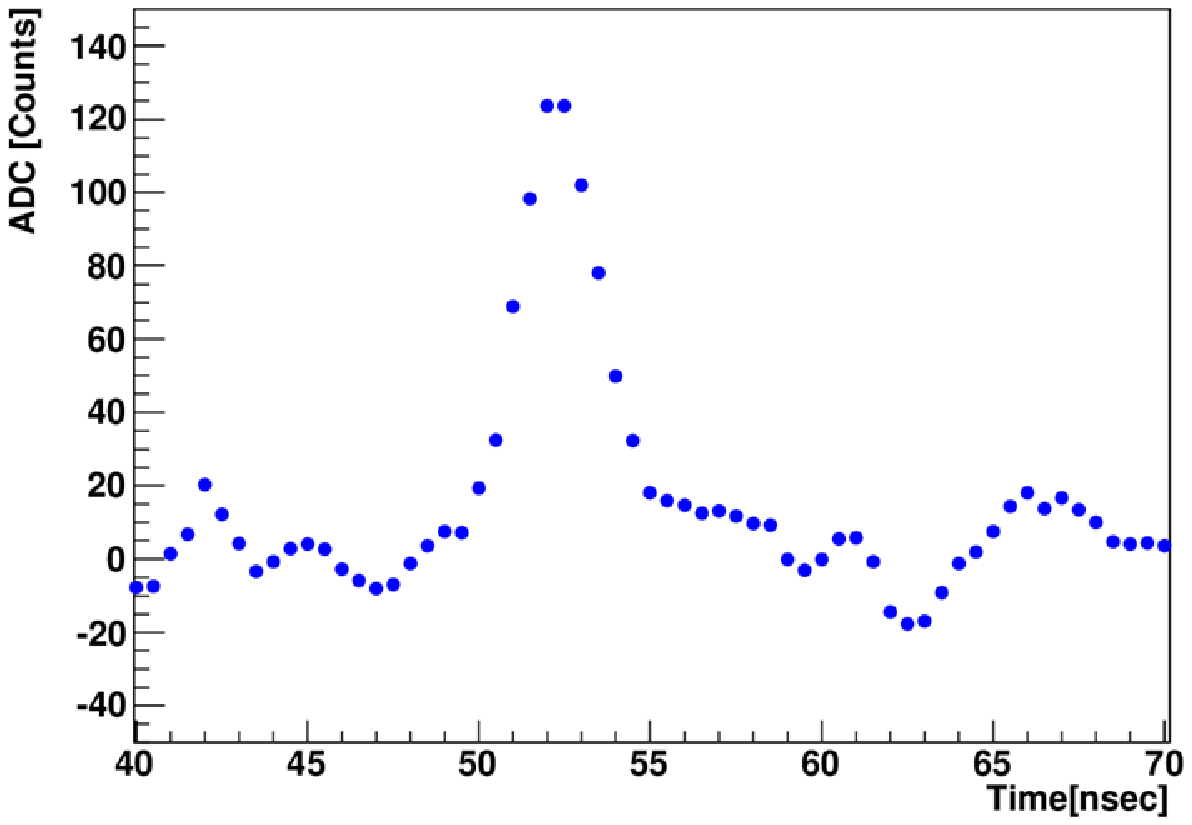}
  \vspace{-3mm}
  \caption{Pulse shape of the PMT signal measured with the DRS4 readout system at a sampling rate of 2 GS/s.}
  \label{fig9}
  \includegraphics[width=5.5cm]{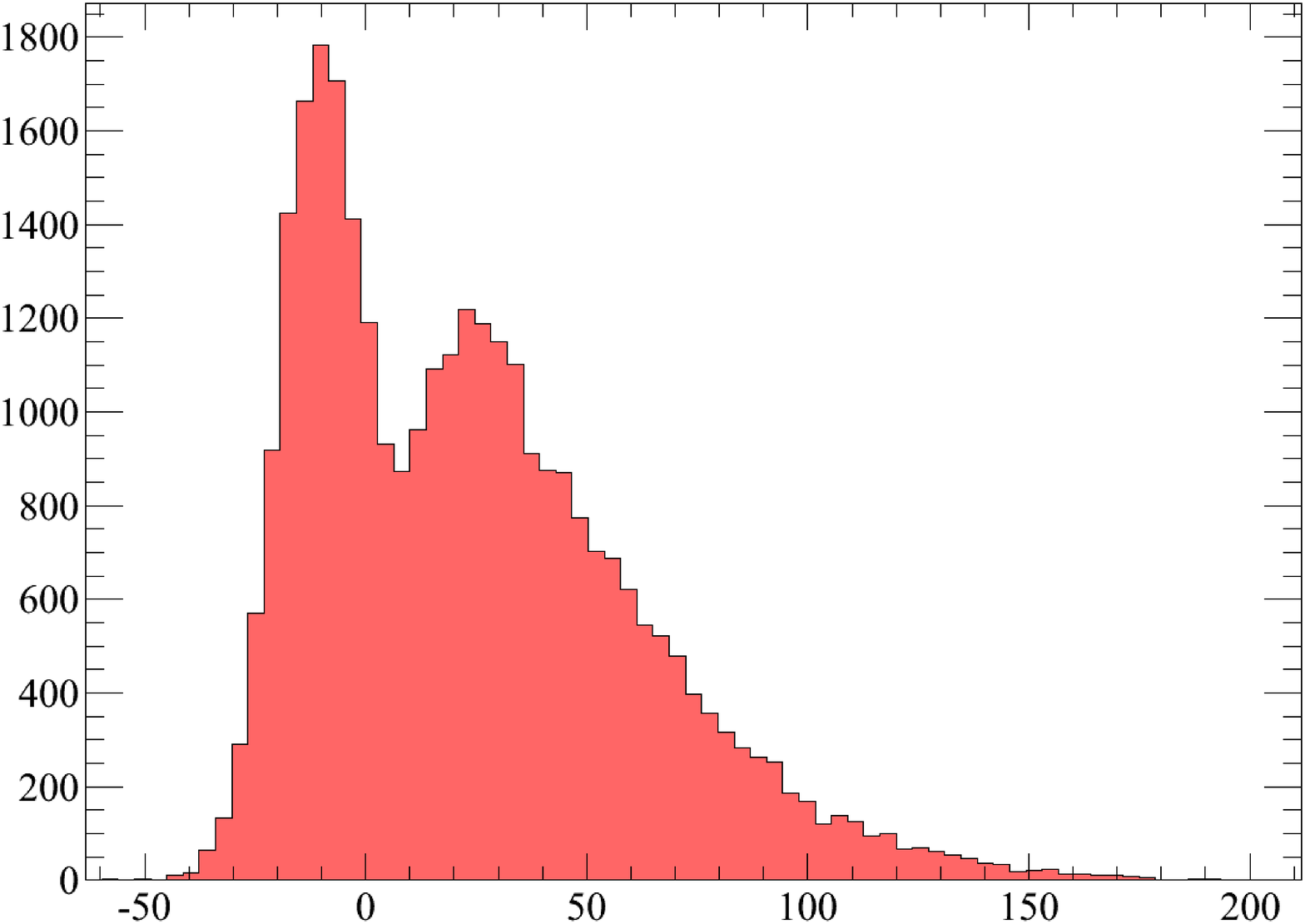}
  \vspace{-5mm}
  \caption{Single photoelectron spectrum of the PMT signal measured with the DRS4 readout system. The horizontal axis is in units of mV $\times$ ns.}
  \label{fig10}
 \end{figure}

\section{Conclusion}
We have developed a prototype 7-PMT cluster readout system for the
large-size telescope of the next generation VHE gamma ray observatory,
CTA. In the readout system named {\it Dragon}, a PMT signal is
amplified, and its waveform is then digitized at a sampling rate of 
the order of GHz using an analog memory ASIC DRS4 that has 4,096
capacitors per readout channel. 
Using the prototype system attached to a PMT with a Cockcroft-Walton
circuit, we successfully obtained a pulse shape of the signal of 
the PMT detecting a UV light at a sampling rate of 2 GS/s, 
and also a single photoelectron spectrum. Evaluation of the readout 
system's performance is ongoing, and a prototype with several dozen 
clusters and their readout boards will be constructed as the \mbox{next} step 
for the development of a full telescope camera.

\vspace*{0.5cm}
\footnotesize{{\bf Acknowledgment:}{We gratefully acknowledge support from the agencies and organisations listed in this page: http://www.cta-observatory.org/?q=node/22}}

\end{document}